\documentclass[12pt]{iopart}

\usepackage{graphicx}
\usepackage[breaklinks=true,colorlinks=true,linkcolor=blue,urlcolor=blue,citecolor=blue]{hyperref}
\usepackage[T1]{fontenc}
\usepackage{ae,aecompl}
\usepackage[dvipsnames]{xcolor}
\usepackage{changes} 

\usepackage{siunitx,braket,multicol,multirow,booktabs}

\usepackage{iopams}
\expandafter\let\csname equation*\endcsname\relax
\expandafter\let\csname endequation*\endcsname\relax
\usepackage{amsmath}

\newcommand{\derivative}{\mathrm{d}}

\DeclareSIUnit\erg{erg}
\DeclareSIUnit\au{au}

\begin{document}

\title[The luminosity constraint in the era of precision solar physics]{The luminosity constraint in the era of precision solar physics}

\author{
Diego Vescovi,$^{1,2,3}$\footnote{These authors contributed equally to this work.}
Carlo Mascaretti,$^{1}$\footnotemark[1]
Francesco Vissani,$^{1,4}$
Luciano Piersanti,$^{2,3}$
and Oscar Straniero$^{3,4}$
}

\address{$^{1}$Gran Sasso Science Institute (GSSI), Viale F. Crispi 7, 67100 L'Aquila, Italy}
\address{$^{2}$INFN, Sezione di Perugia, Via A. Pascoli snc, 06123 Perugia, Italy}
\address{$^{3}$INAF, Osservatorio Astronomico d'Abruzzo, Via Mentore Maggini snc, 64100 Teramo, Italy}
\address{$^{4}$Laboratori Nazionali del Gran Sasso (LNGS), Via G. Acitelli 22, 67100 Assergi (L'Aquila), Italy}

\ead{diego.vescovi@gssi.it}

\begin{abstract}
The \textit{luminosity constraint} is a very precise relationship linking the power released by the Sun as photons and the solar neutrino fluxes. 
Such a relation, which is a direct 
consequence of the physical processes controlling the production and the transport of energy in the solar interior, 
is of great importance for the studies of solar neutrinos 
and has a special role for the search of neutrinos from the CNO cycle, whose first detection with a 5$\sigma$ significance 
has been recently 
announced by the Borexino collaboration. 
Here we revise the luminosity constraint, 
discussing and validating its underlying hypotheses, in the light of latest solar neutrino  and luminosity measurements.
We generalize the current formulation of the luminosity constraint relation so that it can be easily used in future analysis of solar neutrino data, and 
we provide a specific application showing the link between CNO and pp neutrino fluxes.
\end{abstract}

\vspace{2pc}
\noindent{\it Keywords}: Neutrinos -- Nuclear reactions, nucleosynthesis, abundances -- Sun: abundances -- Sun: interior

\submitto{\jpg}

\section{Introduction}\label{sec:intro}

\subsection{Context and motivations}
The Sun, being the closest star, is a fundamental benchmark for our comprehension of the stellar physics and for the stellar evolution theory. The prediction of solar models that the
CNO cycle actually contributes to the energy production in the Sun has never been observationally verified up to now \cite{Agostini:2020mfq}.
Although this contribution is small if compared to the total energy budget, its detection
directly probes the physical conditions within the solar core. Hopefully, it may also help in solving the longstanding problem of the discrepancy between standard solar models (SSMs) and helioseismic measurements. 
In brief, the sound speed profile predicted by SSMs, as computed by using the most recent photospheric abundance determination of CNO and other volatile elements \cite{aspl05,aspl09,caff11}, 
are in tension with the sound speed profile obtained from helioseismic data \cite{basu04}.
In particular, the sound-speed profile allows a precise determination of the location of the internal border of the solar convective envelope and this location is more external in standard solar models than 
the one derived from helioseismic measurement. This issue could be solved if the actual metallicity of the Sun would be larger than currently assumed. Indeed, the location of the convective boundary depends on the temperature gradient and, in turn on the radiative opacity. A larger opacity, as due to a higher metallicity, would imply a deeper convective envelope \cite{chri09,vill10}.
This represents the so-called ``solar metallicity problem'' \cite{bahc}.
It is worth recalling that the solar system abundances currently used in solar model calculations are obtained by combining different sources, mainly spectroscopy of the solar photosphere and mass spectroscopy of pristine meteorites \cite{gs98, aspl05, plj}. Note that this provides us only abundance ratios, rather than absolute abundances \cite{psc}.  
In this context, an independent evaluation of the solar metallicity or of the abundances of its major constituents, i.e. C, N and O, 
through the measurement of the CNO burning rate, may provide this solution.
Note that in this way we may evaluate the abundances of C, N and O in the \textit{solar core}, rather than in the solar photosphere, as obtained from the standard spectroscopic abundance analysis

The CNO burning rate can be determined by measuring the neutrinos related to the $\beta$-decay of $\mathrm{{^{13}N}}$, $\mathrm{{^{15}O}}$, and $\mathrm{{^{17}F}}$. Those are produced in the innermost zone of the Sun, where the temperature is high enough to fully activate the CN cycle and partially the NO cycle. 
The importance of these measurements in our understanding of the present and the primordial Sun has been pointed out by \cite{haxt08} (see also \cite{Zhang_2019} for a recent discussion). In particular, \cite{haxt08} show that the correlation between the core metallicity and the CNO fluxes is independent of other solar model input parameters at a high level of significance and use this information to infer the primordial metallicity of the Sun, provided that a reliable estimation of the diffusion coefficient is available \cite{dals20}.
On an experimental point of view, one has to keep in mind that CNO neutrinos have so low energies ($\sim$ MeV) that their detection is particularly difficult. Moreover, in the same energy region, neutrinos from the pp-chain are also emitted, thus requiring an independent determination of neutrinos from the $\mathrm{p+p\rightarrow {^{2}H}+\e^{+}+\nu}$ and $\mathrm{p+p+{e^-}\rightarrow {^{2}H}+\nu}$ reactions. As a consequence, only two of the solar neutrino telescopes currently working can be used to this aim, namely SAGE in Russia \cite{sage} and Borexino in Italy~\cite{bx}.   
The former provides information about the integrated neutrino flux for energies above \SI{233}{keV}:  in this case the CNO contribution to the total is smaller than the uncertainties in the detection cross section of the apparatus, thus the SAGE determination can be hardly used to constraint the CNO cycle rate.
On the other hand, the detection cross section in Borexino is well known and, in addition, the contamination of the signal from non-CNO neutrinos reduces to the pep reaction only, whose contribution can not be measured directly, thus representing a limitation for any solar neutrino telescope evaluation of the CNO cycle rate.

However, an independent estimate of pep neutrinos can be derived from the solar luminosity, by using a reliable (appropriately simplified) description of the solar structure at the current epoch, based on the assumption, already proved, that the main energy source in the Sun is provided by nuclear reactions \cite{bx}. According to this general scenario, the neutrinos emitted from the Sun are tightly related to its surface luminosity, by means of the so-called \textit{luminosity constraint} whose current formulation is illustrated and discussed in \cite{lum}.
The relevance of such relation relies on the possibility to express the solar photon luminosity, which is measured very precisely, as linear combination of the neutrino fluxes, so that the latter can be linked each other with very high precision.

\subsection{This paper}
In the present work we review such a relation, by critically analyzing the uncertainties on the underlying assumptions and quantifying their contributions to the corresponding estimation of pp and pep neutrino fluxes. 
This kind of analysis has to be performed since neutrino measurements have now attained so high accuracy to require appropriate and very precise theoretical analysis tools.
In this regard we recall that the luminosity constraint is already used for a model-independent analysis of neutrino data \cite{conch}. 
Moreover, recently the Borexino collaboration has announced officially the first detection with 5$\sigma$ confidence level of CNO neutrinos \cite{Agostini:2020mfq}, so it is of paramount importance to improve as possible the luminosity constraint and to establish to what extent it can be used to  estimate pp and pep neutrino fluxes.
Furthermore, it has recently been observed that there are inaccuracies in the formula currently in use for the luminosity constraint \cite{procvissa}, which adds motivation to engage a new discussion of this topic.

The paper is organized as follows:
in Section~\ref{sec:neutrino} we summarize the state-of-the-art of observational results, focusing on the determination of solar neutrino fluxes;  
in Section~\ref{sec:lumcons} we introduce the luminosity constraint in its simplest, ``standard'', form, clarifying the underlying hypotheses and its derivation, and we update the input 
data by introducing their most recent and accurate determination, in particular the solar luminosity value; 
in Section~\ref{sec:generalization} we analyze critically the luminosity constraint and we obtain a more generalized formulation, by releasing the assumption that $^{3}$He and $^{14}$N are in nuclear equilibrium in the Sun and including explicitly the energy contribution coming from solar expansion/contraction;  
in Section~\ref{sec:appli} we illustrate the importance of this result by considering its specific application to the search for CNO neutrinos.
In Section~\ref{s5} we compare our results with the extant literature and summarize our conclusions.

\begin{table}[t!]
\caption{\label{tab:sol_nu_exps}The predicted and measured solar neutrino fluxes
in units of $10^{\gamma_i}$
cm$^{-2}$ s$^{-1}$. The CNO neutrino flux is the sum of the singular fluxes of N, O and F.
Theoretical SSM predictions are from this work (see Sec.~\ref{sec:generalization}), for the GS98 \cite{gs98} and PLJ14 \cite{plj} solar compositions. 
The results from Borexino (BX) are taken from \cite{borex_phaseII,Agostini:2020mfq}, that from Super-Kamiokande (SK) is taken from \cite{sk}, and that for the Sudbury Neutrino Observatory (SNO) is from \cite{sno_boron}. 
We note that SNO has performed a notable oscillation-independent measurement of the Boron flux via the neutral current interaction channel $\nu_\ell + d \to n + p + \nu_\ell$: they obtained $\Phi_\text{B}=5.25(1 \pm 0.04)$, which is compatible with the measurement by Super-Kamiokande.
The error on $\Phi_\text{hep}$ as measured by Super-Kamiokande is not known, but it is presumably large.
In a very recent work by SNO \cite{new_sno}, they quote $\Phi_\text{hep}=(5.1-23)\times \SI{e3}{\per\square\centi\meter\per\second}$ at $1\sigma$.}
\begin{indented}
\item[]\begin{tabular}{@{}lccccc}
\br
\multirow{2}{*}{Flux} & \multirow{2}{*}{$\gamma_i$} &\multicolumn{2}{c}{$\varphi_i$} & \multirow{2}{*}{Experimental results} & \multirow{2}{*}{Source} \\

\cmidrule{3-4} 
 & & GS98 & PLJ14 & & \\
\mr
$\Phi_\text{pp}$ & 10 & $5.99(1 \pm 0.01)$ & 6.01$(1 \pm 0.01)$ &$6.1 (1 \pm 0.1)$ &BX \\
$\Phi_\text{pep}$ & 8 & $1.42(1 \pm 0.02)$ & $1.43(1 \pm 0.02)$ &$1.27(1 \pm 0.17)^{\rm a}$ &BX  \\
$\Phi_\text{Be}$ & 9 & $4.73(1 \pm 0.12)$ & $4.52(1 \pm 0.12)$ &$4.99(1 \pm 0.03)$ &BX \\
$\Phi_\text{B}$ & 6 & $5.52(1 \pm 0.24)$ & $5.01(1 \pm 0.24)$ &$5.41(1 \pm 0.016)$ &SK$^{\rm b}$ \\
$\Phi_\text{hep}$ & 3 & $8.15(1 \pm 0.30)$ & $8.28(1 \pm 0.30)$ &$8   (1 \pm 2)$ &SNO \\
$\Phi_\text{N}$ & 8 & $2.87(1 \pm 0.30)$ & $2.58(1 \pm 0.29)$ & - & - \\
$\Phi_\text{O}$ & 8 & $2.13(1 \pm 0.36)$ & $1.86(1 \pm 0.35)$ & - & - \\
$\Phi_\text{F}$ & 6 & $5.51(1 \pm 0.37)$ & $4.04(1 \pm 0.36)$ & - & - \\
\mr
$\Phi_\text{CNO}$ & 8 & $5.06(1 \pm 0.32)$ & $4.48(1 \pm 0.31)$ &$7.0^{+3.0}_{-2.0}$ &BX  \\
\br
\end{tabular}
\item[] $^{\rm a}$ Measured $\Phi_\text{pep}$ neutrino flux for the GS98 solar composition.
\item[] $^{\rm b}$ By including the SNO measurement mentioned in the caption, $\Phi_\text{B}=5.39 (1\pm 0.015)$.
\end{indented}
\end{table}
%

\section{Experimental status of solar neutrinos}\label{sec:neutrino}

The solar neutrino flux has been studied ever since the Homestake experiment \cite{Cleveland:1998nv}, which has taken data between 1970 and 1994.
Such experiment was based on the inverse $\beta$ absorption of an electron neutrino by chlorine, namely:
$\nu_e + \rm{^{37}Cl} \to \rm{^{37}Ar} +  e^-$.
Today, the most popular techniques of solar neutrino detection rely on Cherenkov emission of neutrino-induced charged particles and neutrino-induced scintillation (see, for reference, \cite{Alimonti_2009,Fukuda:2002uc,Boger_2000}).
Generally, solar neutrino fluxes are expressed as:
\begin{equation}
\Phi_i = \varphi_i \times 10^{\gamma_i}\; \si{\per\square\centi\meter\per\second} \, ,
\label{eq:nufluxes}
\end{equation}
where the fluxes are labeled with $i=\text{pp}$, pep, Be, B, hep, N, O, F, according to their production mechanism. 
Not surprisingly, the $\gamma_i$ exponents have been stable over time, while the adimensional $\varphi_i$ factors have been improved and tested with experiments.

In Table~\ref{tab:sol_nu_exps} we compare the latest experimental determination of solar neutrino fluxes to the theoretical expectations based on SSMs, computed by adopting two 
different heavy metals distributions, namely GS98 \cite{gs98} (high metallicity case) and PLJ14 \cite{plj} (low metallicity case)\footnote{For details about the derivation of the theoretical uncertainties, see Section~\ref{sec:forma}.}.
From Table~\ref{tab:sol_nu_exps} we can see that $\Phi_\text{Be}$ and $\Phi_\text{B}$ are measured with an accuracy better than that of the corresponding theoretical prediction.
This fact, along with the upcoming first measurement of CNO neutrinos by Borexino, marks the beginning of precision solar neutrino flux measurements.
The extraction of the CNO neutrino signal, however, is particularly difficult due to the presence of the background constituted by the decay of bismuth and by pep neutrinos \cite{vill11}.
A precise determination of the bismuth contamination is possible, as it decays $\beta^-$ in polonium in about 20 minutes, which is easily visible and measurable \cite{vill11}.
On the other hand, the identification and separation of pep neutrinos represent a problem not only for Borexino but also for any other neutrino telescope. 
The precise observational knowledge of the solar luminosity can be used to overcome this last experimental difficulty. 
In fact, the solar luminosity can be used to constrain a linear combination of pp and CNO neutrino fluxes, which basically amounts to a precise determination of the pp flux (pep neutrinos are very closely related to pp neutrinos). 
We will discuss both points quantitatively later.
%

\section{The standard luminosity constraint}
\label{sec:lumcons}
\begin{figure}
\centering\includegraphics[width=.85\textwidth]{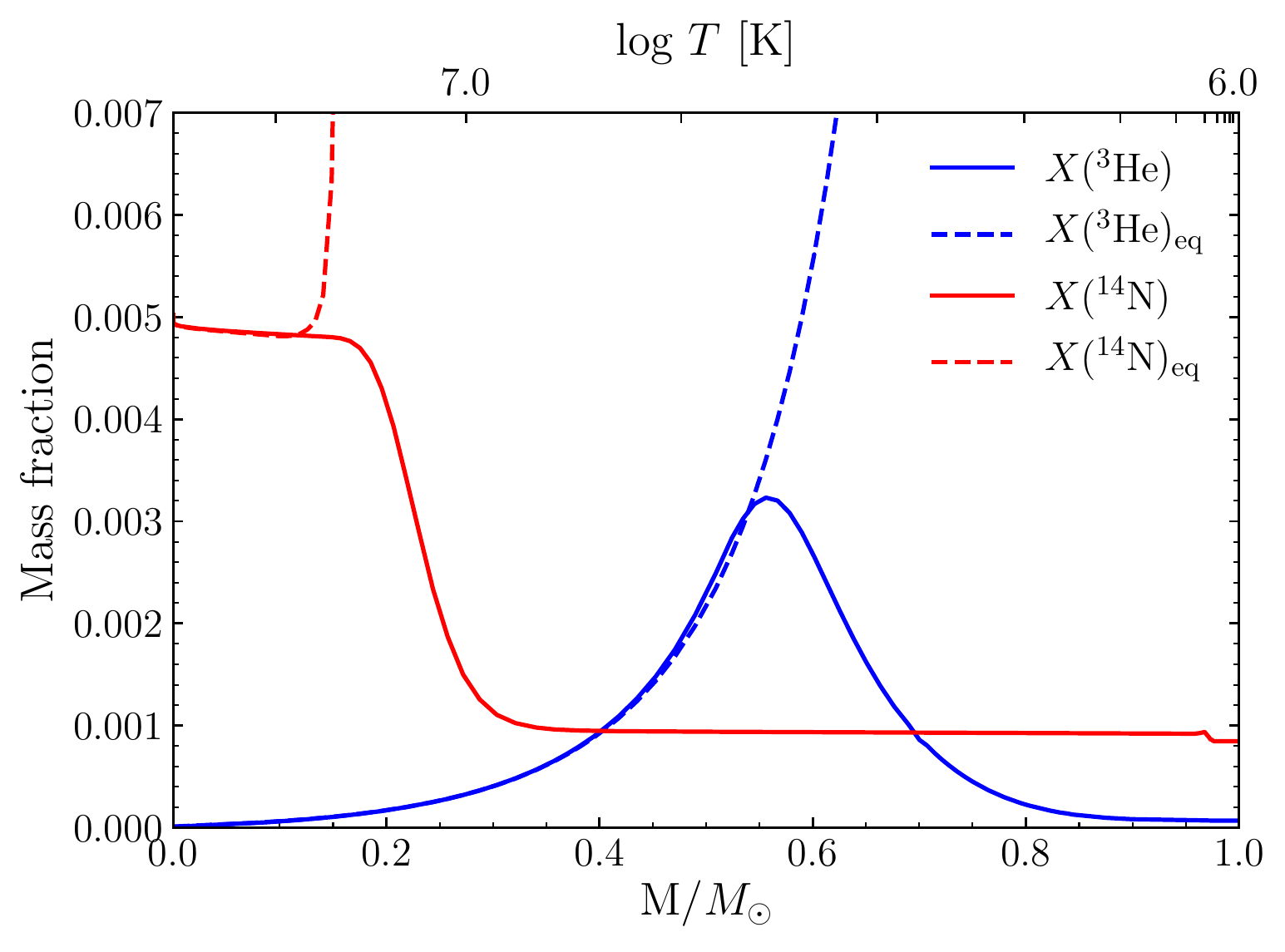}
\caption{\label{fig:equi}$^{3}$He and $^{14}$N abundances (solid lines) as a function of mass coordinate, for the GS98 SSM described in Sec.~\ref{sec:generalization}. 
In the innermost region, production and destruction channels, both for $^{3}$He and $^{14}$N, equate and these isotopes attain their equilibrium abundances (dashed lines). At $\simeq 0.15 M_\odot$ p-captures on ${\rm {}^{12}C}$ and ${\rm {}^{13}C }$ becomes ineffective due to the low temperatures.
$^{3}$He shows a sharp peak at about $0.55M_\odot$: in this region $^{3}$He is continually produced by proton burning reactions but the temperature is too low to burn it at equilibrium rate via ${\rm {}^{3}He + {}^{3}He}$ and ${\rm {}^{3}He + {}^{4}He}$ reactions. In the outer region proton burning is ineffective in producing ${\rm {}^{3}He}$.}
\end{figure}

The luminosity constraint is a relation linking the photospheric solar luminosity with the neutrino fluxes produced in nuclear reactions active in the innermost zones of the Sun, 
and it is obtained under some assumptions representing an excellent approximation of the real solar physical properties \cite{lum}.
First of all, it is assumed  that all the secondary isotopes involved in nuclear processes are in local nuclear equilibrium, \textit{i.e.} that their abundances are fixed by 
the condition that their production rate is equal to the destruction one. Under this hypothesis, 
the net result of these nuclear processes is the conversion of four protons into one $\alpha$ particle mainly via the pp chain, to a small extent via the CNO I cycle, and, to a negligible 
extent, via the CNO II cycle. Every time four protons are destroyed (and a $\mathrm{^{4}He}$ synthesized), two neutrinos are produced.

\subsection{A critical examination of the standard luminosity constraint}

However, the local nuclear equilibrium is exactly verified only in the innermost zones of the Sun, where the temperature is large enough that the production and destruction rates of intermediate isotopes attain nuclear equilibrium.
\begin{figure}
\includegraphics[width=\textwidth]{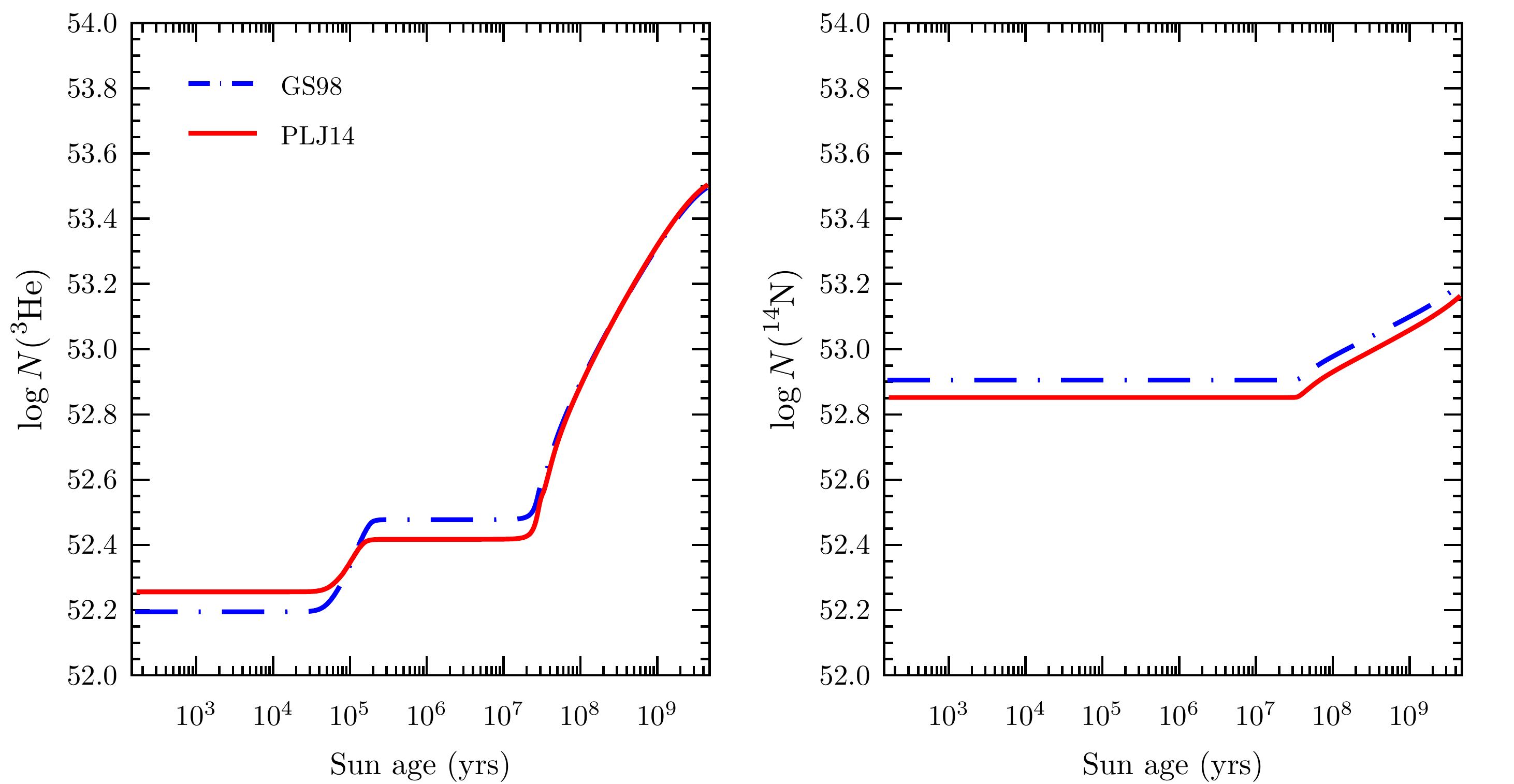}
\caption{\label{fig:temp_evolution}Temporal evolution of the total $^{3}$He (left) and $^{14}$N (right) abundances, adopting both GS98 (blue, dash-dotted line) and PLJ14 (red, solid line) solar compositions.}
\end{figure}
For instance, in the standard formulation of the luminosity constraint \cite{bahc96,lum}, it is assumed that $\mathrm{^{3}He}$ is in local nuclear equilibrium in the whole Sun.
As shown in Fig.~\ref{fig:equi} the equilibrium abundance of $\mathrm{^{3}He}$ (dashed blue line) changes along the structure as the temperature and density values change.
In particular, the $^{3}$He lifetime in the solar core is $\gtrsim 10^{5}$ yr, but it rapidly increases outward, resulting in an exponential increase of the equilibrium abundance. 
On the other hand in the real Sun, due to the temperature decrease, the cross section for $\mathrm{^{3}He}$ production becomes lower and lower for mass coordinate above 
$M_r\sim 0.5-0.6 M_{\odot}$, thus its mass abundance rapidly goes to zero (solid blue line), largely departing from its local equilibrium abundance.

Similar considerations are valid also for the $\mathrm{^{14}N}$: in the innermost zones of the Sun, below $M_r\sim 0.12 M_\odot$, its abundance is determined by the 
efficiency of the CNO cycle, while above this zone, where temperature is lower than $\sim 10^7$ K, it is not produced at all as $p$-captures can occur efficiently on neither 
$\mathrm{^{12}C}$ nor $\mathrm{^{16}O}$ (solid red line in Fig.~\ref{fig:equi}). At variance, the $\mathrm{^{14}N}$ equilibrium abundance in this zone rapidly diverges (dashed red 
line). As a matter of fact, the total number of both $^{3}$He and $^{14}$N nuclei in the whole Sun is expected to increase with the time, as currently obtained by computing SSMs 
(see Fig.~\ref{fig:temp_evolution}).

The other important assumption in deriving the luminosity constraint is that the release of energy in the Sun is due to nuclear reactions only. 
In the Sun (and more in general in real stars during the Main Sequence phase), however, part of the nuclear energy, the so-called ``gravothermal energy'',  is absorbed to produce an expansion against gravity 
on a timescale of few Myr. As a consequence, the surface solar luminosity is expected to be smaller than the energy produced in the central core by nuclear reactions per unit of mass.
Relying on SSM calculations, \cite{bahc96} argued that the gravothermal energy contribution to the solar photospheric luminosity is $\simeq 0.03\%$ , substantially lower than the uncertainty of $0.4\%$ for the value of $L_\odot$ = \SI{3.844e33}{\erg\per\second} \cite{bahc95}, available at that time. 
However, more recent determinations \cite{iau2015} are 10 times more precise and provide for the photospheric solar luminosity the value
\begin{equation}
L_\odot = 3.8275\,(1 \pm 0.0004)\times\SI{e33}{\erg\per\second} ,
\label{eq:solar_lum_phot}    
\end{equation}
therefore, with an estimated uncertainty comparable with the gravothermal energy contribution to the solar luminosity. 

We want to remark that usually in the computation of SSMs no simplifying assumptions are adopted, thus all the relevant physical and nuclear processes are consistently accounted 
for. The results of these computations, however, are based on
 many physical inputs (\textit{e.g.} nuclear cross section, heavy elements initial abundances, radiative opacities, equation of state), simplifying 
hypotheses (\textit{e.g.} approximate description of the convection, neglecting of mass loss processes, assumptions on primordial composition whose potential relevance was recently pointed out \cite{Zhang_2019}), and physical processes, such as gravitational settling of helium and heavy elements, whose uncertainties are definitively larger than those discussed above 
(for a review see \cite{bahc95}). For this reason, the luminosity constraint offers a unique tool to perform a model-independent analysis of the solar neutrino fluxes.

\subsection{A derivation of the standard luminosity constraint}
Before obtaining a more general formulation of the luminosity constraint, accounting for the small deviation from the local equilibrium assumption and for the gravothermal contribution 
to the surface solar luminosity, it is useful to begin by re-deriving this relation in its current ``standard'' form. To this aim we make use of just three assumptions\footnote{As discussed above, the last 2 hypotheses have only approximate validity and will be improved later.}, namely:
\begin{enumerate}
    \item lepton number conservation: the nuclear fusion processes operating in the Sun inevitably produce neutrinos, which can be used to observe and ``tag'' such reactions;
            \item source of the energy: the energy that goes into radiation (neutrinos and photons) comes entirely from nuclear reactions;
                \item net effect of nucleosynthesis; hydrogen nuclei transform into $^{4}$He while all other nuclear species remain unchanged.
\end{enumerate}

The amount of energy delivered from the conversion of four protons into an $\alpha$ particle as a whole is equal to 
\begin{equation} \label{eq:q4}
Q_4:=M_4-4M_1=\SI{26.73097}{\mega\electronvolt} , 
\end{equation} 
where $M_1$ and $M_4$ are the atomic mass of ${^1}$H and ${^{4}}$He, respectively. In this way we automatically take into account the
contribution due to the annihilation of the two positrons from $\beta$-decays with two electrons in the surrounding. 
Such energy eventually goes into photons and neutrinos, so that the solar
luminosity (the power) can be expressed as
\begin{equation} 
L_\text{nuc} = L_\odot + L_\nu .
\label{lumtot}
\end{equation}
Here $L_\odot$ is the solar luminosity emitted in photons, $L_\text{nuc}$ the contribution from nuclear reactions and $L_\nu$ is that emitted in neutrinos, i.e.
\begin{equation}\label{lumnu}
L_\nu =4\SI{\pi}{\square\au} \sum_i \braket{E_i}\Phi_i ,
\end{equation}
where $\braket{E_i}$ is the average energy of neutrinos resulting from a certain reaction and ``au'' is the average Earth-Sun distance.

Considering $^{4}$He as the sole synthesized nucleus, the conservation of energy relation reads
\begin{equation} \label{consenergy}
L_\odot + L_\nu = Q_4\dot N (\rm{^{4}He})  \, ,
\end{equation}
where $N \rm{(^{4}He)}$ is the total number of $^{4}$He nuclei in the Sun and $\dot N \rm{(^{4}He)}$ is its time derivative.\\
As the synthesis of $^{4}$He from $^{1}$H requires the transformation of four protons into two protons and two neutrons, two neutrinos must be produced in the ensuing 
$\beta^+$ processes. We can then write the lepton number conservation as
\begin{equation}
4\SI{\pi}{\square\au}\sum_{i} \Phi_i = 2\dot N (\rm{^{4}He}) 
\label{consleptnumb} ,
\end{equation}
as the rate of neutrino production is twice that of $^{4}$He production.
Solving Eq.~\eqref{consleptnumb} for $\dot N (\rm{^{4}He})$ and substituting this expression and Eq.~\eqref{lumnu} in Eq.~\eqref{consenergy}, we obtain
\begin{equation}
L_\odot + \SI{4\pi}{\square\au} \sum_i \Braket{E_i}\Phi_i = \frac{Q_4}{2}\,\SI{4\pi}{\square\au} \sum_i \Phi_i  \, , 
\end{equation}
thus the luminosity constraint relation becomes
\begin{equation}
\frac{L_\odot}{\SI{4\pi}{\square\au}}=
\sum_i \left(\frac{Q_4}{2}-\Braket{E_i} \right)\Phi_i \, .
\label{lumconstrbahc}
\end{equation}

Please, note that only now, when substituting the numerical values of $Q_4$, $\Braket{E_i}$ (on which we assume negligible uncertainty) as well as choosing which ``kind'' of 
neutrinos are produced, the nuclear physics details are needed.
Throughout this work we adopt the values for $\Braket{E_i}$ provided in Table~2 of \cite{procvissa}.
As an example, let us assume that only neutrinos from the pp-chain are produced in the Sun: then, we would have
$\SI{13.10}{\mega\electronvolt} \times \Phi_\text{pp}=L_\odot / \SI{4\pi}{\square\au}$, and so
$\Phi_\text{pp} = 6.485\,(1\pm 0.0004) \times \SI{ e10}{\per\square\cm\per\second} $, 
the relative error descending directly from that on the solar luminosity value.

Including all the other solar neutrino components 
and dividing both sides of Eq.~\eqref{lumconstrbahc} by $L_\odot/( \SI{4\pi}{\square\au})=
\SI{8.4946}   \times \SI{ e11}{\mega\electronvolt\per\square\cm\per\second} $,
we obtain
\begin{equation} (1\pm 0.04\%)  =\frac{1}{\num{8.4946 e11}} \sum_i k_i \varphi_i \, ,
\label{eq:temp_sol_const} 
\end{equation}
where the $k_i$ coefficients are defined as
\begin{equation}
k_i = \left(\frac{Q_4}{2}-\braket{E_i}\right)
10^{\gamma_i} \label{eq:ki_factors} \, .
\end{equation} 
The values of the $k_i$ coefficients
are listed in Table~\ref{tab:ki_coefficients}, while the $\gamma_i$ factors, as defined in Eq.~\eqref{eq:nufluxes}, can be seen in Table~\ref{tab:sol_nu_exps}.
\begin{center}
\begin{table}
\caption{\label{tab:ki_coefficients}The $k_i$ factors, in units of $\SI{}{\mega\electronvolt\per\square\cm\per\second}$, as defined in Eq.~\eqref{eq:ki_factors}.}
\centering\begin{tabular}{@{}lc}
\br
$i$ & $k_i$ \\
\mr
pp & \num{1.3099e11} \\
Be & \num{1.2552e10} \\
pep & \num{1.1920e9} \\
B & \num{6.6305e6} \\
hep & \num{3.7355e3} \\
N & \num{1.2658e9} \\
O & \num{1.2368e9} \\
F & \num{1.2365e7} \\

\br
\end{tabular}
\end{table}
\end{center}

From the argument above it follows that the luminosity constraint effectively links, within a very small 0.04\% uncertainty, the pp, Be, pep, N and O neutrinos.

\section{A new version of the luminosity constraint}\label{sec:generalization}

The expression derived for the luminosity constraint can be generalized by relaxing the assumption that only ${^{4}}$He is produced in the Sun and accounting for the deviation of ${^{3}}$He and ${^{14}}$N abundances from their equilibrium values in the region above the burning zone. This can be done 
by adding in Eq.~\eqref{consenergy} new terms and weighting them accordingly to the corresponding produced neutrinos. 
In the same way, it is possible also to include energy term related to non-nuclear processes, such as the gravothermal energy. 
The resulting formulation of the luminosity constraint will be more refined and, most importantly, more accurate.

\subsection{The departure from the local nuclear equilibrium}\label{sec:forma}
On a general ground, without any assumptions about the synthesized nuclei, the lepton number conservation as formulated in Eq.~\eqref{consleptnumb} can be expressed as
\begin{equation} 
\SI{4\pi}{\square\au} \sum_i \Phi_i =  \sum_j c_j \dot N(j) \, , 
\label{relationnu}
\end{equation}
where $c_j$ counts the number of electron neutrinos involved in the production of the nucleus $j$; for example, as seen before, $c_{^{4}He}=2$.
The energy conservation (Eq.~\eqref{consenergy}) would then become
\begin{equation} 
L_\odot + L_\nu = \sum_j Q_j\dot N(j) \, ,
\label{consenergymod1}
\end{equation}
as now we are accounting for the production of more nuclei.
As before, we can isolate $\dot N \rm{(^{4}He)}$ from Eq.~\eqref{relationnu} and substitute it into Eq.~\eqref{consenergymod1}, so that:
\begin{equation}
\begin{split}
L_\odot + \SI{4\pi}{\square\au} \sum_i \Braket{E_i}\Phi_i = & \frac{Q_4}{2} \left[\SI{4\pi}{\square\au} \sum_i \Phi_i -\sum_{j\neq \rm{^{4}He}} c_j \dot N(j)\right] \\
& +\sum_{j\neq \rm{^{4}He}}Q_j \dot N(j) \, ,
\end{split}  
\end{equation}
which results in
\begin{equation} \label{lumconstrmod}
\frac{1}{\SI{4\pi}{\square\au}} \left(L_\odot + \sum_{j\neq \rm{^{4}He}} L_{j} \right) =
\sum_i \left(\frac{Q_4}{2}-\Braket{E_i}\right)\Phi_i \, ,
\end{equation}
having defined $L_{j}$ as
\begin{equation}\label{pipolon}
L_{j} := \left( \frac{c_j Q_4}{2}-Q_{j} \right)  \dot N(j) . 
\end{equation}
These $L_{j}$ terms represent the corrections to the power, that can be ``tagged'' by neutrinos (\textit{i.e.} the right-hand-side of Eq.~\eqref{lumconstrmod}), due to the production of the intermediate isotope $j$ other than $^4$He.
As it is well known, the pp chain and the CNO cycle 
can be both naturally broken into two branches, each of them leading to 
the production of one intermediate isotopes and one neutrino, so it turns out $c_j=1$. Therefore, the 
sign of the above quantities $L_{j}$ depends on the energy released per neutrino, and, in the assumption that the abundance of the intermediate isotope increase with time 
(\textit{i.e.}, $\dot N(j) >0$), it is positive if $Q_j/c_j$ is less than $Q_4/c_{\rm ^{4}He}$.

\begin{center}
\begin{table}[t!]
\caption{\label{tab:SSMout}$\dot{N}(j)$ (units of 10$^{35}$ s$^{-1}$) and $L_{j}$ (units of 10$^{30}$ erg s$^{-1}$) quantities, with relative uncertainties, involved in the formulation of the revised luminosity constraint, for GS98 ad PLJ14 models.}
\centering 
\begin{tabular}{@{}lccccc}
\br
Models & $\dot{N}({\rm ^{3}He})$ &$\dot{N}({\rm ^{14}N})$ &$L_{\rm ^{3}He}$  &$L_{\rm ^{14}N}$ & $L_{\rm g}$ \\
\mr
GS98 & $3.29(1 \pm 0.07)$ & $2.15(1 \pm 0.13)$ & $3.39(1 \pm 0.07)$ & $0.57(1 \pm 0.13)$ & $1.54(1 \pm 0.04)$ \\
PLJ14 & $3.42(1 \pm 0.07)$ & $2.07(1 \pm 0.13)$ & $3.53(1 \pm 0.07)$ & $0.55(1 \pm 0.13)$ & $1.52(1 \pm 0.04)$ \\
\br
\end{tabular}
\end{table}
\end{center}

The pp chain is broken into two parts: one corresponding to the first two reactions in the ppI chain ($\mathrm{p(p,\beta^+){^2H}(p,\gamma){^3He}}$), leading to the production of ${^{3}}$He, and the other, corresponding to the last reaction in ppI chain ($\mathrm{{^3He}({^3He},2p){^4He})}$ and those in ppII ($\mathrm{{^3He}({^4He},\gamma){^7Be}(e^-,\nu){^7Li}(p,\gamma){^8Be}\rightarrow 2{^4He}}$) 
and ppIII ($\mathrm{{^3He}({^4He},\gamma){^7Be}(p,\gamma){^8B}\rightarrow\beta^++{^8Be}\rightarrow2{^4He}}$) chains, leading to the formation of one ${^4}$He nucleus.
On the other hand, the CNO I cycle is made by two branches as well, the CN and the NO;
with ${^{14}}$N acting as a bottleneck, due to the very low value of the $p$-capture cross section on this isotope as compared to those on C and O isotopes.

Broadly speaking, the time derivative of total number $\dot N(j)$ of a given isotopes $j$ in Eq.~\eqref{pipolon} can be expressed as
\begin{equation}
\dot N(j) = \frac{\derivative N(j)}{\derivative t} = \frac{N_A}{A_j} \int_{0}^{M_\odot} \frac{\derivative X(j)}{\derivative t} dM ,
\end{equation}
where $N_A$ is the Avogadro number and $X(j)$ is mass fraction and $A_j$ the relative atomic mass number of \textit{j}-nucleus and the integration is performed on the whole mass of the Sun. 

We estimated $\dot N\rm{(^{3}He)}$ and  $\dot N\rm{(^{14}N)}$ by computing two different SSMs, adopting both GS98 \cite{gs98} and PLJ14 \cite{plj} solar compositions, with the use of the FUNS evolutionary code \cite{psc,vesc19}. We used the same input physics as in \cite{vesc19}, except for the solar luminosity, for which we adopted the more precise and recent evaluation of \cite{iau2015}, and equation of state, for which we adopted the OPAL EOS2005 \cite{opal}. We also employed the p+p rate from \cite{adelb11} (see also \cite{marc19}).
In Table~\ref{tab:SSMout} we show, for both GS98 and PLJ14 models, the ensuing predictions for $\dot N\rm{(^{3}He)}$ and $\dot N\rm{(^{14}N)}$. 

In order to calculate the contribution to the luminosity constraint due to 
the non-equilibrium burning of $^{3}$He and $^{14}$N
as described by Eq.~\eqref{lumconstrmod}, we need also the energy released in the production of such nuclei, i.e., 
$Q_3:=M_3-3M_1 = \SI{6.936}{\mega\electronvolt}$ and $Q_{14}:=M_{12}+2M_1-M_{14} = \SI{11.710}{\mega\electronvolt}$.
Finally, as the production of $^{3}$He and $^{14}$N is always accompanied by the emission of one electron neutrino, their $c_j$ factors are both equal to~1.
In Table~\ref{tab:SSMout}, we report also the corrective terms to the luminosity constraint due to the production of $^{3}$He, $^{14}$N, calculated from  Eq.~\eqref{pipolon}.
We find that 
these corrections
are important and not negligible, being comparable or even larger than the present uncertainty on the solar luminosity.

\begin{center}
\begin{table}
\caption{\label{tab:uncert}Estimated 1$\sigma$ uncertainties, for the GS98 (PLJ14) model,  in solar \cite{bahc04,bahc06,iau2015} and nuclear physics \cite{adelb11,mart11,b16}, and their influence on $L_j$ predictions, computed from the partial derivatives of Table~\ref{tab:derivatives}.}
\footnotesize
\centering
\begin{tabular}{@{}lccccc}
\br
$\beta_{j}$ & Central value & $\displaystyle\frac{\Delta \beta}{\beta}$(\%) & $\displaystyle\frac{\Delta L_{\rm ^{3}He}}{L_{\rm ^{3}He}}$(\%) & $\displaystyle\frac{\Delta L_{\rm ^{14}N}}{L_{\rm ^{14}N}}$(\%) & $\displaystyle\frac{\Delta L_{\rm g}}{L_{\rm g}}$(\%) \\
\mr
$L_{\odot}$ & $3.8275 \times 10^{33}$ erg s$^{-1}$ & 0.04 & 0.06 (0.07) & 0.008 (0.009) & 0.04 (0.04)\\
Opacity & 1.0 & 2.5 & 1.54 (1.75) &  0.41 (0.36) & 0.75 (1.08) \\
Age & 4.57 Gyr & 0.44 & 0.73 (0.67) & 0.15 (0.15) & 0.15 (0.16) \\
Diffusion & 1.0 & 15.0 & 2.68 (2.66) & 3.17 (3.21) & 1.80 (1.44) \\
$Z/X$ & 0.02292 & 15.0 & 5.04 (5.28) & 12.83 (12.82) & 3.62 (3.20) \\
p+p & 4.01 $\times 10^{-25}$ MeV b & 1.0 & 0.34 (0.43) & 0.09 (0.09) & 0.44 (0.51) \\
$\mathrm{^3 He + ^3 He}$ & 5.21 MeV b & 5.2 & 2.17 (2.16) & 0.25 (0.23) & 0.55 (0.54) \\
$^3 \mathrm{He}+^{4}\mathrm{He}$ & 0.56 MeV b & 5.4 & 1.35 (1.13) & 0.47 (0.44) & 1.13 (1.10) \\
$\mathrm{p +^{7}Be}$ & 21.3 eV b & 4.7 & 0.008 (0.001) & 0.0003 (0.0001) & 0.01 (0.04) \\
$\mathrm{p +^{14}N}$ & 1.59 keV b & 7.5 & 0.46 (0.52) & 0.14 (0.13) & 0.34 (0.30) \\
\br
\end{tabular}
\end{table}
\end{center}
\begin{table}[t!]
\caption{\label{tab:derivatives}Partial derivatives $\lambda_{j,k}$ of relevant quantities involved in the formulation of the revised luminosity constraint, with respect to solar environmental parameters and S-factors. Table entries are the logarithmic partial derivatives $\lambda_{j,k}$ of the defined quantities $L_{j}$ with respect to the indicated solar model parameter $\beta_{k}$, taken from the GS98 (upper row) and PLJ14 (lower row) SSM best values.}
\footnotesize
\centering
\begin{tabular}{@{}lcccccccccc}
\br
 & \multicolumn{10}{c}{$\beta_{k}$} \\
 \cmidrule{2-11} 
Source &$L_\odot$ & Opacity & Age & Diffusion & \textit{Z/X} & $S_{11}$ & $S_{33}$ & $S_{34}$ & $S_{17}$ &  $S_{114}$\\
\mr
\multirow{2}{*}{$L_{\rm ^{3}He}$} & -1.504 & -0.627 & -1.676 & -0.194 & -0.370 & 0.337 & -0.432 & 0.255 & -0.002 & 0.063 \\
& -1.678 & -0.717 & -1.533 & -0.193 & -0.388 & 0.435 & -0.431 & 0.214 & 0.000 & 0.072 \\
\mr
\multirow{2}{*}{$L_{\rm ^{14}N}$} & 0.191 & -0.166 & -0.332 & 2.094 & 0.864 & 0.095 & 0.050 & -0.089 & 0.000 & -0.020 \\
& 0.233 & -0.146 & -0.336 & 2.119 & 0.863 & 0.095 & 0.045 & -0.084 & 0.000 & -0.018 \\
\mr
\multirow{2}{*}{$L_{\rm g}$} & 1.062 & 0.303 & 0.333 & 0.127 & 0.255 & -0.446 & -0.110 & 0.213 & -0.002 & 0.048 \\
& 1.118 & 0.434 & 0.364 & 0.102 & 0.225 & -0.519 & -0.106 & 0.208 & -0.001 & 0.042
\\
\br
\end{tabular}
\end{table}

We then estimated the uncertainty associated with the quantities $L_{j}$ by considering that
their sensitivity
to the input parameters adopted in the computation of SSMs, $\beta_{k},$ can be expressed in terms of the logarithmic partial derivatives, 
$\lambda_{j,k}$ (see e.g. \cite{bahc88}), defined as
\begin{equation}
\lambda_{j,k}=\frac{\partial \ln L_{j}}{\partial \ln \beta_{k}} \, .
\end{equation}
Hence, the total fractional uncertainty would be 
\begin{equation}\label{ljerror}
\frac{\delta L_{j}}{L_{j}}=\left[\sum_{k}\left[\left(1+ \frac{\Delta\beta_{k}}{\beta_{k}}\right)^{\lambda_{j,k}}-1\right]^{2}\right]^{1 / 2} \, ,
\end{equation}
where the contributions from individual uncertainties are quadratically combined.
Standard deviations computed with this method, which relies on the assumption of linear response of solar models to changes in the input parameters, was shown to agree to better than 10\% with those estimated by Monte Carlo simulations \cite{bahc88}. Due to the small uncertainties of the input parameters, in fact, linearity of solar models is usually a very good approximation \cite{vill10b}.

Such input parameters, which are the major nuclear sources of uncertainties in calculating a SSM, and therefore affecting the predicted $L_j$, can be divided into two sets, corresponding to nuclear and ``solar'' $\beta_k$.
The nuclear parameters are the astrophysical $S$-factors for ${\rm p+p\left(S_{11}\right)},\rm{^3 He}+\rm{^3 He}\left(S_{33}\right),\rm{^3He}+\rm{^4He}\left(S_{34}\right)$, $p+\rm{^{7}Be}\left(S_{17}\right)$, and ${\rm p + ^{14}N\left(S_{114}\right)}$ reactions.
The most important ``solar'' uncertainties arise from the measured photon luminosity $L_{\odot}$, the mean radiative opacity, the solar age, the efficiency of He and metals 
gravitational settling, and the surface metallicity relative to the hydrogen abundance ($(Z/X)_\odot$).
In order to estimate the percentage uncertainties ${\delta L_j}/L_j$ in Eq.~\eqref{ljerror}, we compute many SSMs by varying all the previous listed quantity inside their 
$1 \sigma$ fractional uncertainties $\Delta \beta_{k} / \beta_{k}$ as given in Tab.~\ref{tab:uncert}. In the same Table, we also report the values of the fractional uncertainties on $L_j$, produced by a variation of the individual input $\beta_k$.
Such errors are computed by means of $\lambda_{j,k}$ values listed in Table~\ref{tab:derivatives}.

Eventually, Table~\ref{tab:SSMout} summarizes our best estimates of $L_j$ and their relative theoretical errors.

The correction to the solar luminosity due to the non-equilibrium burning of $^{14}$N can also be evaluated according to the connection among $\dot N \rm{(^{14}N)}$, $\Phi_{\text O}$ and $\Phi_{\text N}$, by assuming that the $^{12}$C $\to$ $^{14}$N and $^{15}$O $\to$ $^{12}$C branches of the CNO I cycle proceed fast (see \textit{e.g.} \cite{procvissa}).
In fact, the emission of a ``nitrogen neutrino'' in the reaction $^{13}$N(e$^+\,\nu_e$)$^{13}$C precedes the production of $^{14}$N, while, every time that a $^{14}$N 
nucleus is destroyed, an ``oxygen neutrino'' is emitted in the reaction $^{15}$O(e$^+\,\nu_e$)$^{15}$N.
This means that the difference in the neutrino fluxes from $\Phi_{\text N}-\Phi_{\text O}$ keeps track of the accumulation of $^{14}$N as
\begin{equation}
\dot N \rm{(^{14}N)} \simeq \SI{4\pi}{\square\au} (\Phi_{\text N}-\Phi_{\text O}) .
\end{equation}
Adopting $\Phi_{\text N}$ and $\Phi_{\text O}$ from Table~\ref{tab:sol_nu_exps}, this results in $\dot N \rm{(^{14}N)} \simeq \SI{2.09e35}{\per\second}$ and $\dot N \rm{(^{14}N)}\simeq \SI{2.02e35}{\per\second}$ for the GS98 and the PLJ14 models, respectively.
The difference with respect to the values reported in Table~\ref{tab:SSMout} arises from a (small) non-zero value for $\dot N \rm{(^{13}C)}$. Note that, in principle, $\dot N \rm{(^{14}N)}$ can be determined from experimental measured values for $\Phi_{\text N}$ and $\Phi_{\text O}$, if available, or alternatively from $\Phi_{\text{CN}} = \Phi_{\text N} + \Phi_{\text O}$, given the stringent value for the $\Phi_{\text N}  / \Phi_{\text O}$ ratio (see Section~\ref{sec:appli}).

\subsection{Non-stationary luminosity constraint}
Due to hydrogen burning, the average molecular weight in the innermost zone of the Sun progressively increases. As a consequence, the solar structure must continually re-adjust 
on a new equilibrium configuration to preserve the hydrostatic equilibrium. This implies that the Sun steadily increases its inner temperature and its photospheric luminosity, 
getting prepared gradually to become a red giant.
In order to account for these changes in the solar structure, the contribution of gravothermal energy has to be included in Eq.~\eqref{consenergymod1} as:
\begin{equation}
L_\odot + L_\nu = \sum_j Q_j\dot N(j) - L_{\rm g} \, .
\label{consenergymod2}
\end{equation}
For a star evolving in time, $L_{\rm g}$ is defined as
\begin{equation}\label{lg_def}
L_{\rm g} \doteq \int_{0}^{M_\odot} \left[ - \frac{\mathrm{d} U}{\mathrm{d} t}+\frac{P}{\rho^{2}} \frac{\mathrm{d} \rho}{\mathrm{d} t}\right] dM \, ,
\end{equation}
where $U$ (erg $\mathrm{g}^{-1}$) is the local internal energy, $P$ (dyne $\mathrm{cm}^{-2}$) is the local pressure, and $\rho$ ($\mathrm{g}$ $\mathrm{cm}^{-3}$) 
is the local density. By adopting this formulation, we are neglecting the energy contribution due to the chemical potential of ions produced/destroyed in nuclear processes because 
this term is order of magnitudes lower than those related to the variation of the internal energy and to the pressure work. 
The quantity $L_{\rm g}$ represents the rate at which the changes in internal energy $\dot{E}_{\rm int} = \int_{0}^{M_\odot} \mathrm{d} U / \mathrm{d} t \, \mathrm{d}M$ and compressional work $\dot{E}_{\rm work} = \int_{0}^{M_\odot} \left(P / \rho^{2}\right)$ $(\mathrm{d} \rho / \mathrm{d} t) \, \mathrm{d}M$ contribute to the stellar luminosity.
The gravothermal luminosity $L_{\rm g}$ can be directly evaluated by approximating the derivative in Eq.~\eqref{lg_def} with finite difference between two models, one at the epoch 
$t=t_\odot$ (see Tab.~\ref{tab:uncert}) and the other at the epoch $t=t_\odot - \mathrm{d}t$, where $\mathrm{d}t$ is the last evolutionary time step.
We computed $L_{\rm g}$ both for GS98 and PLJ14 models (see Table~\ref{tab:SSMout}). Analogously to Section~\ref{sec:generalization}, we estimated the global uncertainty 
affecting $L_{\rm g}$, by considering all the SSM-inputs with their errors, and computing the relative partial derivatives. Tab.~\ref{tab:derivatives} shows the partial derivatives 
$\lambda_{j,k}$.
We remark once again that the estimated values of $L_{\rm g}$  are comparable with the present uncertainty for $L_\odot$ and, therefore, should be included for a precise 
derivation of the luminosity constraint.

\subsection{Summary}\label{summary} 
Basing on Eq.~\eqref{consenergymod2} and Eq.~\eqref{lumconstrmod} we can now write the luminosity constraint in its general form as 
\begin{equation}\label{lumconstrmod2}
\frac{1}{\SI{4\pi}{\square\au}} \left(L_\odot + L_{\rm ^{3}He} + L_{\rm ^{14}N}+ L_{\mathrm{g}} \right) =
\sum_i \left(\frac{Q_4}{2}-\Braket{E_i}\right)\Phi_i \, . 
\end{equation}
The values of the new quantities that appear in the equation above are given in Table~\ref{tab:SSMout},
for the two SSMs considered in this work.
We note that the first two terms  $L_{\rm ^{3}He}$ and $L_{\rm ^{14}N}$ are both positive (see Section~\ref{sec:forma}), the latter being particularly small
because the energy released in the production of $^{14}$N is only a bit less than $Q_4/2$.
The rate of gravothermal energy production $L_{\rm g}$, determined by the rate of change of the molecular weight in the innermost zones of the Sun, is gained at the expense 
of the energy delivered by nuclear reactions: 
as a consequence, the same amount of nuclear reactions (and of neutrinos) would correspond to a photon luminosity diminished by $L_{\mathrm{g}}$.
As the solar luminosity is fixed by observations, and the three new terms are all positive, this means that their net effect is to increase the number of expected neutrinos.
This is quite evident by putting Eq.~\eqref{lumconstrmod2} in the same form of Eq.~\eqref{eq:temp_sol_const}, namely 
\begin{equation} (1 \pm \sigma)  =\frac{1}{\mathcal F} \sum_i k_i \varphi_i \, ,
\label{eq:temp_sol_const2}
\end{equation}
where the values of the $k_i$ are still the one given in Table~\ref{tab:ki_coefficients}, and
\begin{equation}
\mathcal F = \frac{L_\odot+L_{^3\rm{He}}+L_{^{14}\rm{N}}+L_\text{g}} {4\pi\,\si{\square\au}} \end{equation} 
where 
$$\mathcal F(\text{GS98}) = 8.5068\times 10^{11} \qquad \mathcal F(\text{PLJ14}) = 8.5070 \times 10^{11} $$
The error $\sigma$ in the previous equation can be evaluated with simple error propagation from Table \ref{tab:SSMout}: in both GS98 and PLJ14 composition models we obtain $\sigma = 0.04\%.$
Summarizing the result of this discussion, the new contributions can be included by changing the value of the coefficient in the denominator of Eq.~\eqref{eq:temp_sol_const} with the new coefficient $\mathcal F$.

\section{Luminosity constraint and the search for CNO neutrinos}
\label{sec:appli}

In view of the experimental state-of-the-art of solar neutrino measurements, it is 
interesting to illustrate the use of the  luminosity constraint to improve the search for CNO neutrinos.

From Table~\ref{tab:ki_coefficients} it is clear that neutrinos from CNO, in particular N and O neutrinos, play an important role in the determination of $\varphi_\text{pp}$ in the context of the luminosity constraint, which helps linking $\varphi_\text{pp}$ and $\varphi_\text{CNO}$ to each other.
In fact, the luminosity constraint by itself is not enough to improve our knowledge on these neutrino fluxes; however, when 
further theoretical information on the $\Phi_\text{pep}/\Phi_\text{pp}$ and on the $\Phi_\text{O}/\Phi_\text{N}$ ratios is 
provided, then a stringent relation between the pp and CNO fluxes can be obtained.
At first glance, one could believe that the usage of this theoretical information would make this application model-dependent, but this is not the case.

The ratio of the pp and pep reaction rates is fixed by nuclear physics \cite{adelb11}, and such rate is assumed also for the pp and pep neutrino fluxes which, 
\textit{a priori}, could be different due to the impact of the whole ensemble of processes occurring in the Sun. 

Comparing how the pep to pp neutrino fluxes change with different metallicities, we get:
$
\left.\Phi_\text{pep} / \Phi_\text{pp} \right|_\text{GS98}=\num{2.375e-3} (1\pm 0.012)$
and 
$
\left.\Phi_\text{pep} / \Phi_\text{pp} \right|_\text{PLJ14}=\num{2.383e-3} (1\pm 0.012)$
The uncertainty of each ratio was obtained 
combined quadratically the errors as discussed in  Section~\ref{sec:forma}. 
This result reinforces the assumption that such a ratio is largely model-independent and that it depends on nuclear physics only. 

Regarding the ratio of N and O neutrinos, if we assume that only $^{4}$He is produced, its value is exactly 1, due to the nuclear equilibrium of all nuclei in the CNO cycles.
As discussed in Section~\ref{summary}, however, because of the slowness ${\rm {}^{14}N(p,\gamma){}^{15}O}$, the CN cycle does not reach the equilibrium and ${}^{13}{\rm N}$ neutrinos are slightly more abundant than the ${}^{15}{\rm O}$ neutrinos.
The ratio $\Phi_\text{O}/\Phi_\text{N}$ is then fixed to be $\text{<}$ 1, with a value depending upon the adopted SSM. 
For the two SSMs we calculated, the ratios are:
$ \left.\Phi_\text{O} / \Phi_\text{N}\right|_\text{GS98}= 0.742 (1\pm 0.053) $
and
$\left.\Phi_\text{O} / \Phi_\text{N}\right|_\text{PLJ14}= 0.722 (1\pm 0.056) $.
Such a variation due to different SSMs affects the luminosity constraint so slightly (less than 0.02\%) that the model-dependence is not a significant issue also in this regard.

The same procedure can be used to link the F neutrinos to the N ones; in this case
$ \left.\Phi_\text{F} / \Phi_\text{N}\right|_\text{GS98}= 0.019 (1\pm 0.085) $ and $ \left.\Phi_\text{F} / \Phi_\text{N}\right|_\text{PLJ14}= 0.016 (1\pm 0.086) $. 
Thanks to the arguments above, 
the required ratios of neutrino fluxes are  
\begin{align}\label{apini1}
&\Phi_\text{pep}/\Phi_\text{pp}=
2.379 (1 \pm 0.012)\times 10^{-3}\\\label{apini2} &\Phi_\text{O}/\Phi_\text{N}=0.732 (1 \pm 0.06)\\
&\Phi_\text{F}/\Phi_\text{N}=0.017 (1 \pm 0.12)\label{apini3} \, .
\end{align}
Such values are obtained by averaging the two results and keeping into account the theoretical errors and the difference between the two predictions, summing in quadrature these uncertainties.  
Note that the first uncertainty is similar, slightly smaller, to the value 1.4\% adopted by Borexino collaboration \cite{agostini2020a,Agostini:2020mfq}.

Now, we proceed to derive of a constraint between the pp and N neutrino fluxes using these ratios and including the information on the fluxes which have been already measured.
After Borexino phase-II  \cite{borex_phaseII}, $^{7}$Be neutrinos are very well known 
and their flux is fixed with a precision better than the theoretical one to
\begin{equation}
\Phi_\text{Be}=(4.99 \pm 0.11^{+0.06}_{-0.08})\times \SI{e9}{\per\square\centi\meter\per\second}  \, . 
\end{equation}
Also the boron neutrino flux, after four phases of Super-Kamiokande operation \cite{sk}, is very well known to be
\begin{equation}
\Phi_\text{B} =5.41(1 \pm 0.016) \times \SI{e6}{\per\square\centi\meter\per\second} \, .
\end{equation}
Note that, while the value of the beryllium flux is quite important for the luminosity constraint, the value of the boron flux has a very small relevance; therefore, the inclusion of the SNO measurement, given in Tab.~\eqref{tab:sol_nu_exps}, has no impact in practice.
%

Eq.~\eqref{eq:temp_sol_const} can be rewritten as
\begin{equation}
\begin{split}
(1 \pm 0.04\%) = &\phantom{+}0.15420\, \varphi_\text{pp}+ 0.01478 \,\varphi_\text{Be}+ 0.00140\, \varphi_\text{pep} \\
& + 0.00149\,\varphi_\text{N}+ 0.001456\, \varphi_\text{O} + \num{1.46e-5}\,\varphi_\text{F}\\
&+\num{7.81e-6}\,\varphi_\text{B}.
\end{split}
\end{equation}
Introducing the ratios obtained in Eqs.~\eqref{apini1}-\eqref{apini3}, we have:
\begin{equation}
\begin{split}
(1 \pm 0.04\%)    =&\phantom{+}0.15453\, \varphi_\text{pp}+  0.002556\,\varphi_\text{N}\\
&+0.01478 \,\varphi_\text{Be}+\num{7.81e-6}\,\varphi_\text{B}
\end{split}
\label{eq:p1}
\end{equation}
and now we can subtract the beryllium and boron contribution:
\begin{equation}
(0.9262 \pm 0.0022) = 0.15453\, \varphi_\text{pp}+ 0.002556\,\varphi_\text{N} \label{eq:p2} 
\end{equation}
thus, isolating $\Phi_\text{pp}$:
\begin{equation} \label{eq:lumconstrnew}
\Phi_\text{pp} + 1.654\, \Phi_\text{N}= 5.994\,(1 \pm 0.2\%)\times\SI{ e 10}{\per\square\centi\meter\per\second} 
\end{equation}
The main contribution to the 0.2\% error is due to the experimental uncertainty in the beryllium neutrino flux; therefore, in principle, this can be decreased in the future. 
As already remarked, the contribution of the boron flux is much less relevant.

The impact of the uncertainties on the of ratios Eqs.~\eqref{apini1}-\eqref{apini2} modify the prefactor of $\Phi_\text N$ in Eq.~\eqref{eq:lumconstrnew}, in that its 1$\sigma$ range is $ 1.654(1\pm 0.025) $.
Given the fact that $\Phi_\text{pp}$ is roughly more than 200 times larger than $\Phi_\text N$, 
we conclude that a 2.5\% variation of the $\Phi_\text{N}$ prefactor induces an error of 0.02\%, which 
is negligible for the purposes of  Eq.~\eqref{eq:lumconstrnew}.
We can then conclude that the uncertainties in  Eqs.~\eqref{apini1}-\eqref{apini3} do not introduce any significant model dependence in Eq.~\eqref{eq:lumconstrnew}.

\begin{table}[t!]
\caption{\label{tab:recapcorrections}The central value, in units of $\SI{ e 10}{\per\square\centi\meter\per\second}$, of the constraint (with 0.2\% precision) 
as described in Eq.~\eqref{eq:lumconstrnew}, including the various refinements of Section \ref{sec:generalization}.}\vskip2mm
\centering\begin{tabular}{@{}lccc}
\br
corrective terms & GS98 & PLJ14 &average\\
\mr
none &5.9936 &5.9936 &5.9936 \\
\mr
$L_{\rm ^{3}He}$ &5.9994  &5.9996 &5.9995\\
\mr
$L_{\rm ^{3}He} + L_{\rm ^{14}N}$  &6.0004 &6.0006 &6.0005 \\
\mr
$L_{\rm ^{3}He} + L_{\rm ^{14}N}+ L_{\mathrm{g}}$ &6.0030 &6.0031  &6.0031\\
\br
\end{tabular}
\end{table}
In order to include the corrections described in Section~\ref{sec:generalization}, it is sufficient to replace the right-hand side value of the above equation with the values given 
in Table~\ref{tab:recapcorrections}. 
Including the contribution due to the corrective terms $L_{\rm ^{3}He}$, $L_{\rm ^{14}N}$, and $L_{\mathrm{g}}$ one obtains:
\begin{equation} \label{eq:lumconstrnew_final}
\Phi_\text{pp} + 1.654\, \Phi_\text{N}= 6.003\,(1 \pm 0.2\%)\times\SI{ e 10}{\per\square\centi\meter\per\second} \, .
\end{equation}
Note that, since the relative difference between the corrective terms for GS98 and PLJ14 models is about $\simeq$ 0.003\% (see last row of Table~\ref{tab:recapcorrections}), this relation links pp and nitrogen neutrino fluxes without a significant impact on the assumptions for the solar core metallicity.
Eq.~\eqref{eq:lumconstrnew_final} can also be expressed in terms of $\Phi_\text{CNO} = \Phi_\text{N} + \Phi_\text{O} + \Phi_\text{F}$, as
\begin{equation} \label{eq:lumconstrnew_cno}
\Phi_\text{pp} + 0.946\, \Phi_\text{CNO}= 6.003\,(1 \pm 0.2\%)\times\SI{ e 10}{\per\square\centi\meter\per\second} \, .
\end{equation}

\section{Conclusive remarks}\label{s5}

In this work we derived an improved version the luminosity constraint, by relaxing two of the fundamental assumptions adopted in the original standard derivation of \cite{lum} as well as in \cite{procvissa}, 
\textit{i.e.} the local nuclear equilibrium of all the isotopes involved in the transformations of 4 protons into a ${^{4}}$He and the stationarity of the solar structure.
The relation we obtained represents the most straightforward, complete and useful one currently available.

In his seminal work \cite{lum}, Bahcall stated:
\begin{quote}
If nuclear fusion reactions among light elements are responsible for the solar luminosity, then  a  specific  linear  combination  of  solar  neutrino  fluxes  must  equal the  solar  constant [$\ldots$]
\begin{equation*}
\frac{L_\odot}{4\pi \si{\square\au}} = \sum_i \left(\frac{\alpha_i}{\SI{10}{MeV}}\right) \Phi_i 
\end{equation*}
[$\ldots$] The coefficient $\alpha_i$ is the amount of energy provided to the star by nuclear fusion reactions associated with each of the important solar neutrino fluxes,~$\Phi_i$.
\end{quote}
In our formulation, as detailed in Section~\ref{sec:lumcons}, the factors analogous to Bahcall's $\alpha_i$ are:
\[\alpha_i = 10^{-\gamma_i-1} k_i, \] \,
where the $\gamma_i+1$ comes from the normalization of $\alpha_i$ factors to 10 MeV. For the neutrinos coming from the CNO cycles we obtain in our formalism:
\begin{align}
\alpha_{\text N} &= \frac{M_{12}+M_1-M_{13}- \braket{E_{\text N}}}{\SI{10}{MeV}} =\num{0.34570} \\
\alpha_{\text O} &= \frac{3M_1+M_{13}-M_4-M_{12}-\braket{E_{\text O}}}{\SI{10}{MeV}}= \num{2.157} \\
\alpha_{\text F} &= \frac{M_{16}+M_1-M_{17}-\braket{E_{\text F}}}{\SI{10}{MeV}}=\num{0.2361} 
\end{align}
where we make use of the updates values for the nuclear masses of the involved isotopes.
We want to remark that such an approach does not distinguish the CNO bi-cycle at its slowest node, \textit{i.e.} at ${}^{14}$N, and this represents the main reason for the 
critical analysis performed in \cite{procvissa}.

In Table~\ref{tab:comparabah} we compare our results with those from \cite{lum} and, as it is quite evident, the factors are very close to each others. 
This demonstrates that our approach is fully consistent with that used in \cite{lum}. 
There is, however, a notable exception regarding the $\alpha_\text{Be}$ value. In the computation 
of that factor, Bahcall stated:
\begin{quote}
[$\ldots$] one must average over the two $^{7}$Be neutrino lines with the appropriate weighting \emph{and include the $\gamma$-ray energy from the 10.3\% of the decays that go to the first excited state of $^{7}$Li}.
\end{quote}
We believe that this procedure leads to double counting the largest energy that neutrinos can have in such a decay, as first noted in \cite{procvissa}.

In Eq.~\eqref{eq:lumconstrnew} we provide a form of the luminosity constraint which is ready to be used in pp and CNO neutrino analyses, something impossible 20 years ago, by incorporating the neutrino fluxes from ${^{7}}$Be and ${^{8}}$B, which are currently very well determined experimentally. This was envisioned by Bahcall in \cite{lum}:
\begin{quote}
    In the future, the generalized luminosity constraint can and should be implemented in analyses that determine solar neutrino parameters.  
    The additional constraint provided by the measured solar luminosity will be especially important when pp and $^7$Be neutrino fluxes are measured as well as the $^8$B neutrino flux.  
    As more experimental data become available, the analyses of neutrino oscillations will become more independent of the standard solar model and it will be natural and convenient 
    to incorporate the luminosity constraint.
\end{quote}
\begin{table}
\caption{\label{tab:comparabah}
The matching of Bahcall's formalism from \protect\cite{lum} ($\alpha_i$ column) to ours.}\vskip 2mm
\centering
\begin{tabular}{@{}lccc}
\br
\multirow{2}{*}{$i$} &$\alpha_i$&&$ 10^{-\gamma_i-1}k_i$ \\
& \cite{lum} &&this work \\
\mr
pp &\num{1.30987} &&\num{1.30987} \\ 
pep &\num{1.19193} &&\num{1.19205} \\
hep &\num{0.37370} &&\num{0.37355} \\
Be &\num{1.26008} &&\num{1.25525} \\ 
B &\num{0.66305} &&\num{0.66305} \\ 
\br
\end{tabular}
\end{table}

Our formulation of the luminosity constraint is more accurate, as it includes more precise values of the solar luminosity and of the atomic masses as compared to \cite{lum}. Moreover, it is also more general, as it allow us to include the effects of non-equilibrium in nuclear processes as well as non-nuclear energy terms, as detailed in Eq.~\eqref{lumconstrmod}.

In particular, we explicitly considered the effects of non-equilibrium abundances of $^3$He and $^{14}$N
and the gravothermal energy contribution to the total energy budget, thus improving the constraint on the linear combination of all solar neutrino fluxes (see Eq.~\eqref{eq:temp_sol_const2}). Basing on such equation we derived a ready-to-use relation, linking the pp and CNO neutrino fluxes (see Eq.~\eqref{eq:lumconstrnew_cno}) and we presented in Tab.~\ref{tab:recapcorrections} the impact of the considered corrective terms above.

The very high precision of the measured photospheric solar luminosity (Eq.~\eqref{eq:solar_lum_phot}) gives us the possibility to test the relevance of such corrections, 
even if in the final expression (Eq.~\eqref{eq:lumconstrnew}) such refinements do not have a sizable impact.
This is due to the fact that the current determination of the $^7$Be flux is not sufficiently precise - compare Eqs.~\eqref{eq:p1} and \eqref{eq:p2}. 
This is the limiting factor of the current numerical precision of the luminosity constraint, namely 0.2\%.

\begin{figure}[t!]
\centering
\includegraphics[width=.85\textwidth]{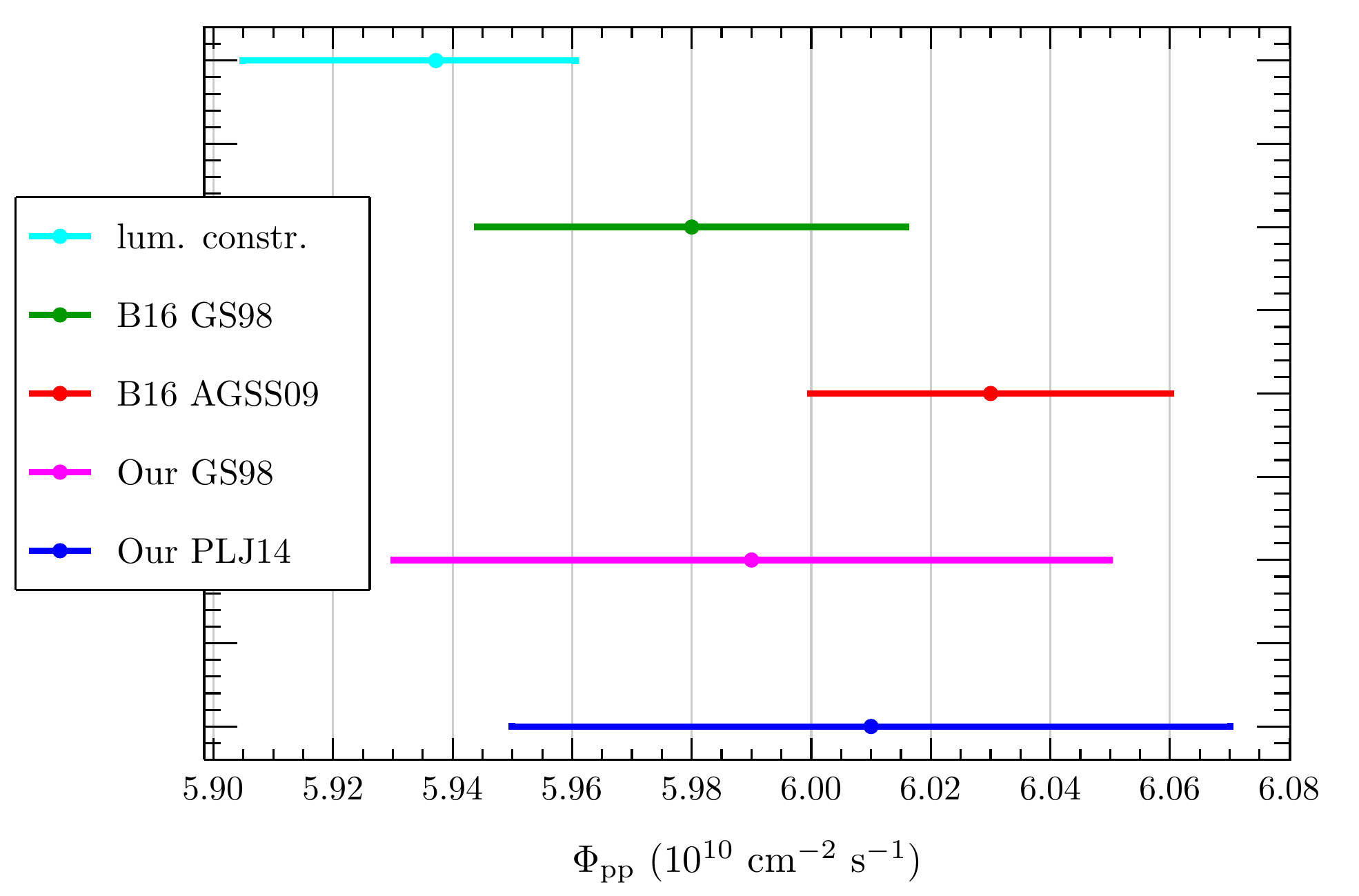}
\caption{\label{fig:confronto_pp} The comparison among the theoretical predictions on the pp neutrino flux from \cite{b16} (``B16''), our theoretical predictions (see Table~\ref{tab:sol_nu_exps}), and the constrained value resulting from incorporating the CNO flux (as measured by Borexino in \cite{Agostini:2020mfq}) in Eq.~\eqref{eq:lumconstrnew_cno}, in which all corrections are considered (blue cyan bar).
The bars show the $1\sigma$ range.}
\end{figure}

As a final application, we make use of 
Eq.~\eqref{eq:lumconstrnew_cno} along with the very recent measurement of CNO neutrinos by Borexino collaboration (see Table~\ref{tab:sol_nu_exps}), 
to derive the corresponding pp neutrino flux:
\begin{equation} \label{eq:deducedppflux}
\Phi_\text{pp} = 5.937_{-0.032}^{+0.023}\times\SI{ e 10}{\per\square\centi\meter\per\second}.
\end{equation}
The asymmetric range reflects the range for the CNO flux found by Borexino. Note that if we used the ``naive'' version of the luminosity constraint (the one that leads to Eq.~\eqref{eq:lumconstrnew} rather than Eq.~\eqref{eq:lumconstrnew_final}) the central value would decrease from $5.937$ to $5.928$, namely by 0.2\%, the same amount of the uncertainty in the luminosity constraint.
Moreover, is the uncertainty in the measurement of the CNO neutrinos that dominates the error in $\Phi_{\mbox{\tiny pp}}$ and not the 0.2\% uncertainty in the luminosity constraint; thus, improved measurements will reduce this error.

The value of the flux in Eq.~\eqref{eq:deducedppflux} is compared in Fig.~\ref{fig:confronto_pp} with several theoretical SSM predictions:  see 
the value indicated with the label \textit{lum.~constr.}. 
Such a value is marginally consistent, within the estimated uncertainties, with theoretical predictions for models assuming an high abundance of heavy elements (\textit{B16 GS98}, \textit{Our GS98}) and, to a lesser extent, for \textit{Our PLJ14} model. A possible explanation for such an evidence is that SSMs currently predict a too low CNO abundance in the solar core. This interpretation is in agreement with the analysis performed in \cite{gough19} who suggested that the total metal content in the solar core ($Z_c$) where CNO neutrinos are produced linearly depends on 
$\Phi_{\rm CNO}$. In fact, by making use of the value measured by the Borexino collaboration for the CNO neutrino flux and the relation provided by \cite{gough19} (see their Eq.~(1)), the estimated metallicity in the solar core should be $Z_c=0.028^{+0.012}_{-0.008}$, whereas for the high metallicity SSM computed in the present work (GS98) we obtain 
$Z_c= 0.02024 \pm 0.00731$.
This fact could give new insights and raise new questions regarding the metal content in solar core 
and/or possibly the approximations adopted in the current versions of the SSM.


\section*{Acknowledgements}
This work was partially supported by the research grant number 2017W4HA7S ``NAT-NET: Neutrino and Astroparticle Theory Network'' under the program PRIN 2017 funded by the Italian Ministero dell'Universit\`a e della Ricerca (MIUR). 

\section*{References}
\bibliographystyle{iopart-num}
\bibliography{biblio}


\end{document}